\documentclass[conference]{IEEEtran}

\usepackage{algorithm}
\usepackage{algorithmicx}
\usepackage{xspace}
\usepackage{amsmath,amssymb}
\usepackage{algpseudocode}
\usepackage[compress,sort]{cite}
\usepackage[pdftex]{graphicx}
\usepackage{url}
\usepackage{subcaption}
\usepackage{color}
\usepackage[keeplastbox]{flushend}

\newcommand{\vct}[1]{\ensuremath{\boldsymbol{#1}}} 
\newcommand{\mat}[1]{\ensuremath{\mathbf{#1}}}
\newcommand{\set}[1]{\ensuremath{\mathcal{#1}}}
\newcommand{\T}{\ensuremath{\top}}

\newcommand{\argmin}{\operatornamewithlimits{\arg\,\min}}

\newcommand{\myparagraph}[1]{\smallskip \noindent \textbf{#1.}}
\newcommand{\ie}{{i.e.}\xspace}
\newcommand{\eg}{{e.g.}\xspace}
\newcommand{\etal}{{et al.}\xspace}

\begin{document}
\title{Adversarial Malware Binaries: Evading Deep Learning for Malware Detection in Executables}

\author{
	\IEEEauthorblockN{
		Bojan Kolosnjaji\IEEEauthorrefmark{1},
		Ambra Demontis\IEEEauthorrefmark{2},
		Battista Biggio\IEEEauthorrefmark{2}\IEEEauthorrefmark{3},
		Davide Maiorca\IEEEauthorrefmark{2},
		Giorgio Giacinto\IEEEauthorrefmark{2}\IEEEauthorrefmark{3},\\
		Claudia Eckert\IEEEauthorrefmark{1},
		and
		Fabio Roli\IEEEauthorrefmark{2}\IEEEauthorrefmark{3}}

	\IEEEauthorblockA{\IEEEauthorrefmark{1}Technical University of Munich\\kolosnjaji@sec.in.tum.de ,claudia.eckert@in.tum.de, \\
	\IEEEauthorrefmark{2}University of Cagliari, Italy\\ \{ambra.demontis, battista.biggio, davide.maiorca, giacinto, roli\}@diee.unica.it \\
	\IEEEauthorrefmark{3}Pluribus One, Italy} }

\maketitle

\begin{abstract}
Machine-learning methods have already been exploited as useful tools for detecting malicious executable files. They leverage data retrieved from malware samples, such as header fields, instruction sequences, or even raw bytes, to learn models that discriminate between benign and malicious software. However, it has also been shown that machine learning and deep neural networks can be fooled by evasion attacks (also referred to as adversarial examples), i.e., small changes to the input data that cause misclassification at test time. In this work, we investigate the vulnerability of malware detection methods that use deep networks to learn from raw bytes. We propose a gradient-based attack that is capable of evading a recently-proposed deep network suited to this purpose by only changing few specific bytes at the end of each malware sample, while preserving its intrusive functionality. Promising results show that our adversarial malware binaries evade the targeted network with high probability, even though less than $1\%$ of their bytes are modified.
\end{abstract}

\IEEEpeerreviewmaketitle

\section{Introduction} 
Detection of malicious binaries still constitutes one of the major quests in computer security~\cite{symantec17}.
To counter their growing number, sophistication and variability, machine learning-based solutions are becoming increasingly adopted also by anti-malware companies~\cite{kaspersky-report}.

Although past research work on binary malware detection has explored the use of traditional learning algorithms on $n$-gram-based, system-call-based, or behavior-based features~\cite{schultz01-sp,assaleh04-sac,rieck08-dimva,yan13_dimva}, more recent work has considered the possibility of using deep-learning algorithms on raw bytes as an effective way to improve accuracy on a wide range of samples~\cite{raff17-arxiv}. The rationale is that such algorithms should automatically learn the relationships among the various sections of the executable file, thus extracting a number of features that correctly represent the role of specific byte groups in specific sections (\eg, if a byte belongs to the code section or simply to a section pointer).  

While machine learning can be used to map the features from malware analysis to a decision on classifying programs as benign or malicious, this process is also vulnerable to adversaries that may manipulate the programs in order to bypass detection. It has been shown that deep-learning methods and neural networks are particularly vulnerable to these evasion attacks, also known as \emph{adversarial examples}, \ie, input samples specifically manipulated to be misclassified~\cite{biggio13-ecml,szegedy14-iclr}. While the existence of adversarial examples has been widely demonstrated on computer-vision tasks (see, \eg, a recent survey on the topic~\cite{biggio18}), it is common to consider that it is not trivial to practically implement the same attack on executable files~\cite{xu16-ndss,anderson17-blackhat,raff17-arxiv}. This is because one mistake at changing the code section or the headers may completely compromise the file functionality. 

In this work, we show that, despite the various challenges required to modify binary sections, it is still possible to compromise the detection of deep-learning systems for malware detection by performing few changes to malware binaries, while preserving their functionality. 
In particular, we introduce a gradient-based attack to generate \emph{adversarial malware binaries}, \ie, evasive variants of malware binaries. The underlying idea of our attack is to manipulate some bytes in each malware to maximally increase the probability that the input sample is classified as benign. Although our attack can ideally manipulate every byte in the file, in this work we only consider the manipulation of padding bytes appended at the end of the file, to guarantee that the intrusive functionality of the malware binary is preserved. We nevertheless discuss throughout the paper which other bytes and  sections of the file can be modified while still preserving its functionality.
Our attack is conceived against \emph{MalConv}, \ie, a deep neural network trained on raw bytes for malware binary detection, recently proposed by Raff~\etal~\cite{raff17-arxiv}.
To our knowledge, this is the first time that such an attack is proposed at the \emph{byte-level} scale, as most work in adversarial machine learning for malware detection has considered injection and removal of API calls or similar characteristics~\cite{yang17-acsac,demontis17-tdsc,grosse17-esorics,biggio13-ecml,maiorca18-arxiv,zhang16-tcyb,chen17-acsac,tong17-arxiv,huang18-arxiv}.

We perform our experiments on 13,195 Windows Portable Executable (PE) samples, showing that the accuracy of \emph{MalConv} is decreased by over 50\% after injecting only $10,000$ padding bytes in each malware sample, \ie, less than 1\% of the bytes passed as input to the deep network.
We also show that our attack outperforms random byte injections, and explain why being capable of manipulating even fewer bytes \emph{within} the file content (rather than appending them at the end) may drastically increase the success of the attack.

\begin{figure*}[t]
\centering
\includegraphics[width=0.85\textwidth]{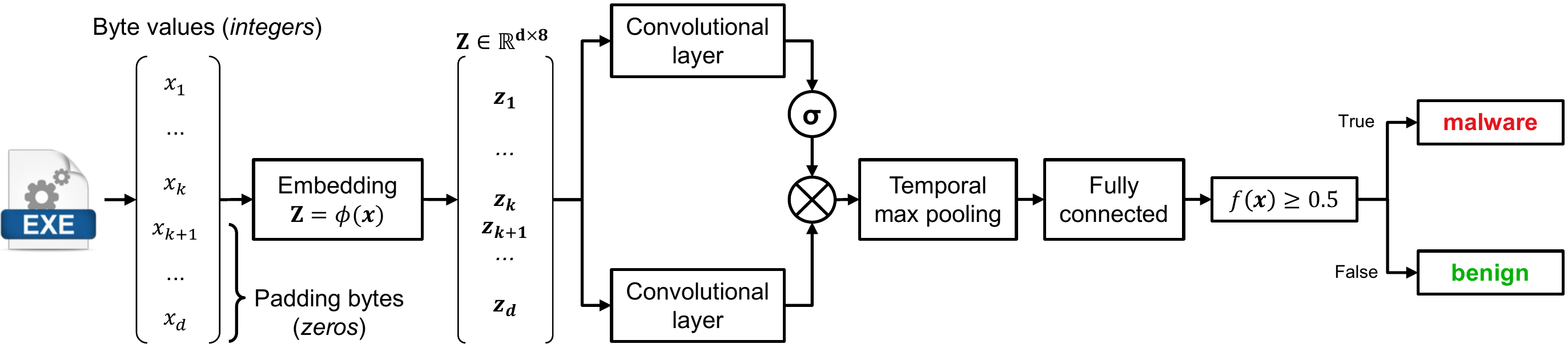}
\vspace{6pt}
 \caption{Architecture of the \emph{MalConv} deep network for malware binary detection~\cite{raff17-arxiv}.}
\label{fig:malconv}
\end{figure*}

With this paper, we aim to claim that it may be very difficult to deploy a robust detection methodology that blindly analyzes the executable bytes. Learning algorithms can not automatically learn the hard-to-manipulate, \emph{invariant} information that reliably characterizes malware, if not proactively designed to keep that into account~\cite{biggio14-tkde}, either by providing proper training examples or encoding a-priori knowledge of which bytes can be maliciously manipulated.
Robustness against adversarial attacks provided by well-motivated miscreants is thus a crucial design characteristic. This work provides preliminary evidence of this issue, which we aim to further investigate in the future.

\section{Portable Executable (PE) Format}
\label{sec:basics}

We provide here a brief description of the structure of PE files, and the prominent approaches that can be used to practically change their bytes.

\subsection{PE File Basics}

PE files are executables that are characterized by an organized structure, which will be briefly described in the following (more details can be found in~\cite{pe-report}).

\myparagraph{Header} A data structure that contains basic information on the executable, such as the number and size of its sections, the operating system type, and the role performed by the file itself (\eg, a dynamically-linked library). Such header is organized in three sub-sections: (i) a DOS header, as the first bytes of a PE executable essentially represent a DOS program; (ii) the true PE header; (iii) an \emph{optional header} which contains information such as the entry point of the file (\eg, the address of the first loaded instruction), the size of the code sections, the magic number, etc.  

\myparagraph{Section Table} A table that describes the characteristics of each file section, with a special focus on a virtual address range that represents how that section will be mapped in memory once the process is loaded. It also contain clear references to where the data generated by the compiler/assembler are stored for each section.

\myparagraph{Data} The actual data related to each section. The most important ones are \texttt{.text} (which contains code instructions), \texttt{.data} (which contains the initialized global and static variables), \texttt{.rdata} (which contains constants and additional directories such as debug), and \texttt{.idata} (which contains information about the used imports in the file).

\subsection{Manipulating PE Files}

Manipulating PE files with the goal of preserving their functionality is in general a non-trivial task, as it can be quite easy to compromise them by even changing one byte. As reported by Anderson~\etal~\cite{anderson17-blackhat}, possible and simple solutions to perform manipulations include either \emph{injecting} bytes in part of the files that are not used (\eg, adding new sections that are never reached by the code), or directly \emph{appending} them at the end of the file. Of course, these strategies are prone to detection by simply inspecting the file header or the section table (in the simplest case of byte appending), or by checking if such sections are accessed by the code itself (in case of more complex injections). 

There are some special cases in which it is possible to directly perform changes to the executable without compromising its functionality. A popular example is changing bytes related to debug information, which are simply used as reference by code developers. Packing (\ie, compressing part of the executable that is then decompressed at runtime) is another possibility, which is however not adequate to perform fine-grained modifications to the file.

More complex changes require precise knowledge of the architecture of the file, and may be not always feasible. For instance, changing the \texttt{.text} section may entirely break the program. However, more trivial changes can be quite dangerous for the file integrity; for example, adding bytes to an existing section would require changing the header and section table accordingly. For the sake of simplicity, in this paper we only refer to byte appending as modification strategy.

\section{Deep Learning for Malware Binary Detection}

The deep neural network attacked in this paper is the \emph{MalConv} network proposed by Raff~\etal~\cite{raff17-arxiv}, depicted in Fig.~\ref{fig:malconv}. 
Let us denote with $\set X = \{0,\ldots,255\}$ the set of possible integer values corresponding to a byte.
Then, the aforementioned network works as follows. The $k$ bytes $(x_1, \ldots, x_k) \in \set X^k$ extracted from the input file are padded with zeros to form an input vector $\vct x$ of $d$ elements (if $k <  d$, otherwise the first $d$ bytes are only considered without padding). This ensures that the input vector provided to the network has a fixed dimensionality regardless of the length of the input file.
Each byte $x_j$ is then embedded as a vector $\vct z_j = \phi(x_j)$ of $8$ elements (through a fixed mapping $\phi$ learned by the network during training). This amounts to encoding $\vct x$ as a matrix $\mat Z \in \mathbb R^{d \times 8}$. This matrix is then fed to two convolutional layers, respectively using Rectified Linear Unit (ReLU) and sigmoidal activation functions, which are subsequently combined through \emph{gating}��~\cite{dauphin16-arxiv}. This mechanism multiplies element-wise the matrices outputted by the two layers, to avoid the vanishing gradient problem caused by sigmoidal activation functions. The obtained values are then fed to a temporal max pooling layer which performs a $1$-dimensional max pooling, followed by a fully-connected layer with ReLU activations. To avoid overfitting, Raff~\etal~\cite{raff17-arxiv} use DeCov regularization~\cite{cogswell15-arxiv}, which encourages a non-redundant data representation by minimizing the cross-covariance of the fully-connected layer outputs. The deep network eventually outputs the probability of $\vct x$ being malware, denoted in the following with $f(\vct x)$. If $f(\vct x) \geq 0.5$, the input file is thus classified as malware (and as benign, otherwise).

\section{Adversarial Malware Binaries}
\label{sec:algorithm}

We discuss here how to manipulate a source malware binary $\vct x_0$ into an \emph{adversarial} malware binary $\vct x$ by appending a set of carefully-selected bytes after the end of file. As in previous work on evasion of machine-learning algorithms~\cite{biggio13-ecml}, our attack aims to minimize the confidence associated to the malicious class (\ie, it maximizes the probability of the adversarial malware sample being classified as benign), under the constraint that $q_{\rm max}$ bytes can be injected.
Note that, to append $q_{\rm max}$ bytes to $\vct x_0$, we have to ensure that $k + q_{\rm max} \leq d$, where $k$ is the size of $\vct x_0$ (\ie, the number of informative bytes it contains) without considering the padding zeros.
This means that the maximum number of bytes that can be injected by the attack is $q = \min (k+q_{\rm max}, d) - k$.\footnote{Note that $q \leq 0$ if $k \geq d$, which means that no byte can be manipulated by this attack.}
This can be characterized as the following constrained optimization problem:
\begin{eqnarray}
\min_{\vct x} && f(\vct x) \, , \\
{\rm s.t.} && d(\vct x, \vct x_0) \leq q \, ,
\end{eqnarray}
where the distance function $d(\vct x, \vct x_0)$ counts the number of padding bytes in $\vct x_0$ that are modified in $\vct x$.

\begin{figure}[t]
    \centering
    \includegraphics[width=0.40\textwidth]{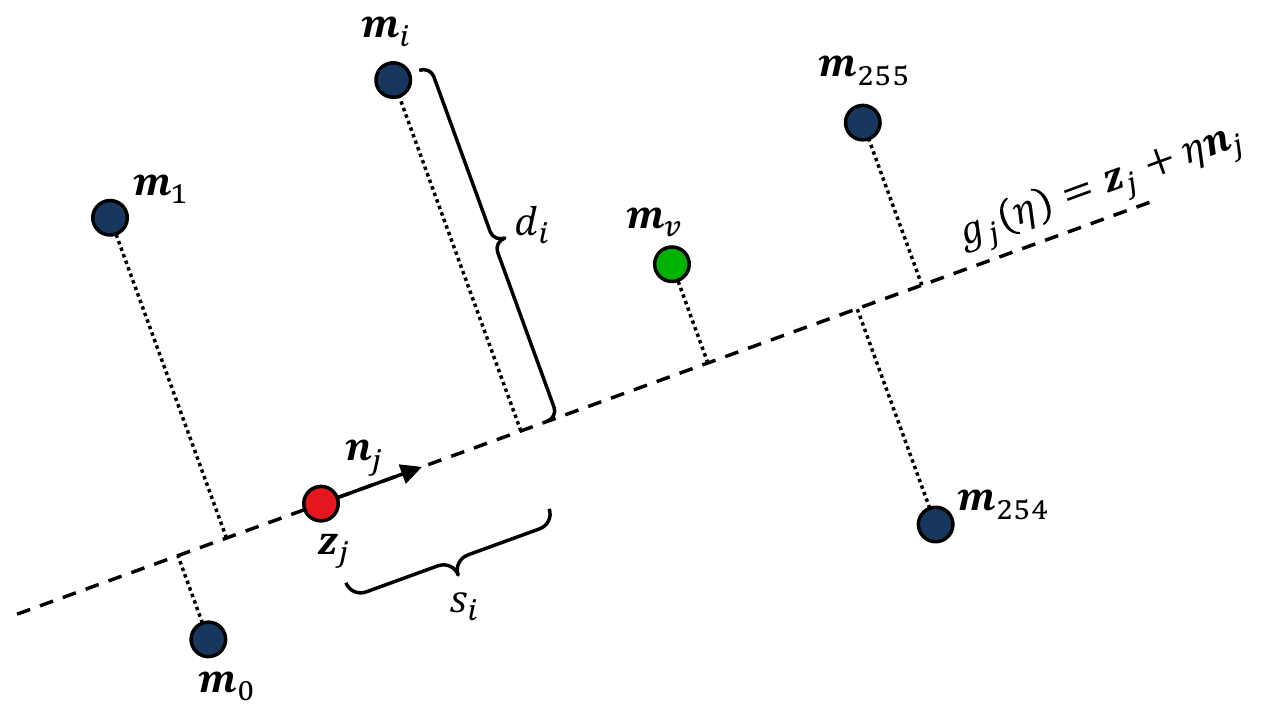}
 \caption{
 Representation of an exemplary two-dimensional byte embedding space, showing the distance $d_i$ and the projection length $s_i$ of each byte $\vct m_i$ with respect to the line $g_j(\eta)$. In this case, the padding byte $\vct z_j$ will be modified by the attack algorithm to $\vct m_v$, as $d_v = \min_{i : s_i > 0} d_i$, \ie, $\vct m_v$ is the closest byte with a projection on $g_j(\eta)$ aligned with $\vct n_j$.}
     \label{fig:projection}
\end{figure}

\begin{algorithm}[t]
	\caption{Adversarial Malware Binaries}
	 \label{alg:attack_algorithm}
	\begin{algorithmic}[1]
		\Require $\vct x_{0}$, the input malware (with $k$ informative bytes, and $d-k$ padding bytes); 
		$q$, the maximum number of padding bytes that can be injected (such that $k+q \leq d$);
		$T$, the maximum number of attack iterations.
		\Ensure $\vct x^{\prime}$: the adversarial malware example.
		\State Set $\vct x = \vct x_0$.
		\State Randomly set the first $q$ padding bytes in $\vct x$.
		\State Initialize the iteration counter $t=0$.
		\Repeat
		\State Increase the iteration counter $t \leftarrow t+1$.
			\For{$p = 1, \ldots, q$}
			\State Set $j=p+k$ to index the padding bytes.
			\State Compute the gradient $\vct w_j = - \nabla_\phi (x_j)$.
			\State Set $\vct n_j = \vct w_j / \| \vct w_j \|_2$.
			\For{$i = 0, \ldots, 255$}
			\State Compute $s_i = \vct n_j^\T (\vct m_i - \vct z_j)$.
			\State Compute $d_i = \|  \vct m_i - (\vct z_j  + s_i \cdot \vct n_j) \|_2$.
			\EndFor
			\State Set $x_j$ to $\argmin_{i : s_i > 0} d_i$.
			\EndFor
		\Until{$f(\vct x) < 0.5$ or $t \geq T$}
		\State \Return $\vct x^\prime$
	\end{algorithmic}
\end{algorithm}

We solve this problem with a gradient-descent algorithm similar to that originally proposed in~\cite{biggio13-ecml}, by optimizing the padding bytes one at a time.
Ideally, we would like to compute the gradient of the objective function $f$ with respect to the padding byte under optimization. However, the \emph{MalConv} architecture is not differentiable in an end-to-end manner, as the embedding layer is essentially a lookup table that maps each input byte $x_j$ to an 8-dimensional vector $\vct z_j = \phi(x_j)$. 
We denote the embedding matrix containing all bytes with $\mat M \in \mathbb R^{256 \times 8}$, where the row $\vct m_i \in \mathbb R^{8}$ represents the embedding of byte $i$, for $i=0,\ldots,255$.
To overcome the non-differentiability issue of the embedding layer, we first compute the (negative) gradient of $f$ (as we aim to minimize its value) with respect to embedded representation $\vct z_j$, denoted with $\vct w_j = - \nabla_\phi(x_j) \in \mathbb R^8$.
We then define a line $g_j(\eta) = \vct z_j + \eta \vct n_j$, where $\vct n_j = \vct w_j / \| \vct w_j \|_2$ is the normalized (negative) gradient direction. This line is parallel to $\vct w_j$ and passes through $\vct z_j$. The parameter $\eta$ characterizes its geometric locus, \ie,
by varying $\eta \in (-\infty, \infty)$ one obtains all the points belonging to this line.
Ideally, assuming that the gradient remains constant, the point $\vct z_j$ will be gradually shifted towards the direction $\vct n_j$ while minimizing $f$. We thus consider a good heuristic to replace the padding byte $x_j$ with that corresponding to the  embedded byte $\vct m_i$ closest to the line $g_j$, provided that its projection on the line is aligned with $\vct n_j$, \ie, that $s_i = \vct n_j^\T (\vct m_i - \vct z_j) > 0$.
Recall that the distance of each embedded byte $\vct m_i$ to the line $g_j$ can be computed as $d_i = \|  \vct m_i - (\vct z_j  + s_i \cdot \vct n_j) \|_2$. A conceptual representation of this discretization process is shown in Fig.~\ref{fig:projection}.
This procedure is then repeated for each modifiable padding byte (starting from a random initialization),
and up to a maximum number of iterations $T$, as described in Algorithm~\ref{alg:attack_algorithm}.

\myparagraph{Generation of Adversarial Malware Binaries} Although the padding bytes are generated by manipulating the input vector $\vct x$, creating the corresponding executable file without corrupting the malicious functionality of the source file is quite easy, as also explained in Sect.~\ref{sec:basics} and in~\cite{anderson17-blackhat}.
It is however worth mentioning that our attack is general, \ie, it can be used to manipulate any byte within the input file. To this end, it suffices to identify which bytes can be manipulated without affecting the file functionality, and optimize them (instead of optimizing only the padding bytes). 

\section{Experiments}

We practically reproduced the deep neural network proposed in \cite{raff17-arxiv}, and performed the evasion attacks according to the algorithm described in Sect.~\ref{sec:algorithm}. In the following, we first describe the employed setup, and then we discuss the results obtained by comparing the efficiency of the proposed gradient-based method with trivial random byte addition.

\myparagraph{Dataset} We employed a dataset composed of $9,195$ malware samples, which were retrieved from a number of sources including \texttt{VirusShare}, \texttt{Citadel} and \texttt{APT1}. Additionally, to evaluate the performances of the network we employed $4,000$ benign samples, randomly retrieved and downloaded from popular search engines. 

\myparagraph{Network Performances} We evaluated the performances of the deep neural network by splitting our dataset into a training and a test set, each of them containing $50\%$ of the samples of the initial dataset. To avoid results that could be biased by a specific training-test division, we repeated this process three times and averaged the results. Under this setting, we obtained an average precision of $92.83\pm5.56$\% and an average recall of $84.68\pm11.71$\% (mean and standard deviation).  

\subsection{Results on Evasion Attacks}

\begin{figure}
  \centering
    \includegraphics[width=0.4\textwidth]{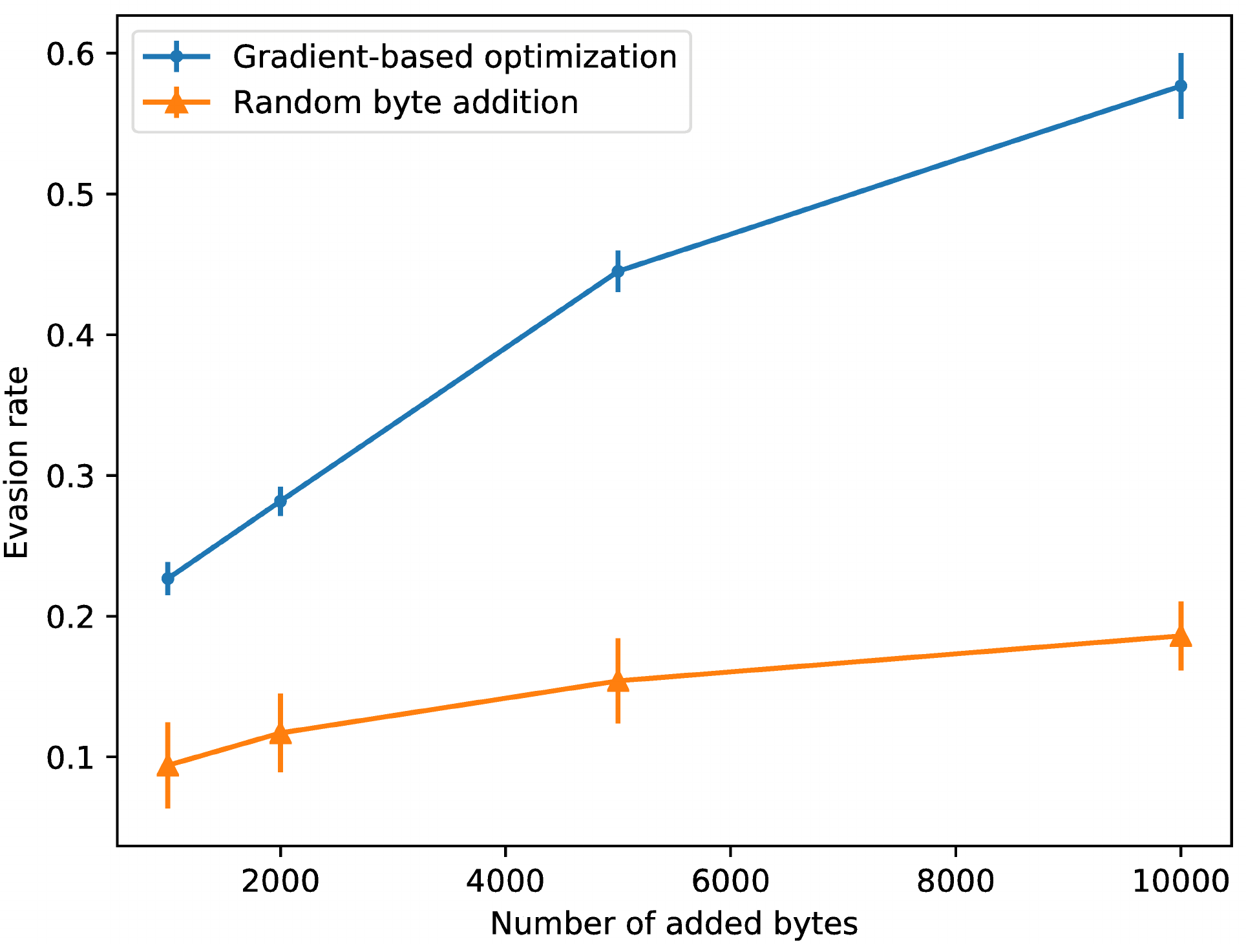}
    \caption{Evasion rate against number of injected bytes.}
    \label{fig_evasion_rate}
\end{figure}

We performed our tests by modifying $200$ randomly-chosen malicious test samples with Algorithm~\ref{alg:attack_algorithm} to generate the corresponding \emph{adversarial malware binaries}. 
As for Algorithm~\ref{alg:attack_algorithm}, we set the maximum number of attack iterations $T=20$, and the maximum number of injected bytes $q_{max}=10,000$. As a result, we chose all malware samples that satisfied the condition $k+q_{max}\leq d$, where $k$ is the file size and $d=10^6$. 
 The attack was performed by appending, at the end of each file, bytes that were chosen according to two different strategies: a \emph{random} attack injecting random byte values, and our \emph{gradient-based} attack strategy.
To verify the efficacy of the attack, we measured for each amount of added bytes the average \emph{evasion rate}, \ie, the percentage of malicious samples that managed to evade the network. 
Fig.~\ref{fig_evasion_rate} provides the attained results as the number of bytes progressively increases, averaged on the three aforementioned training-test splits. Notably, adding random bytes is not really effective to evade the network.
Conversely, our gradient-based attack allows evading \emph{MalConv} in $60\%$ of the cases when $10,000$ padding bytes are modified, even if this amounts to manipulating less than 1\% of the input bytes.

\begin{figure}
\centering
        \includegraphics[width=0.24\textwidth]{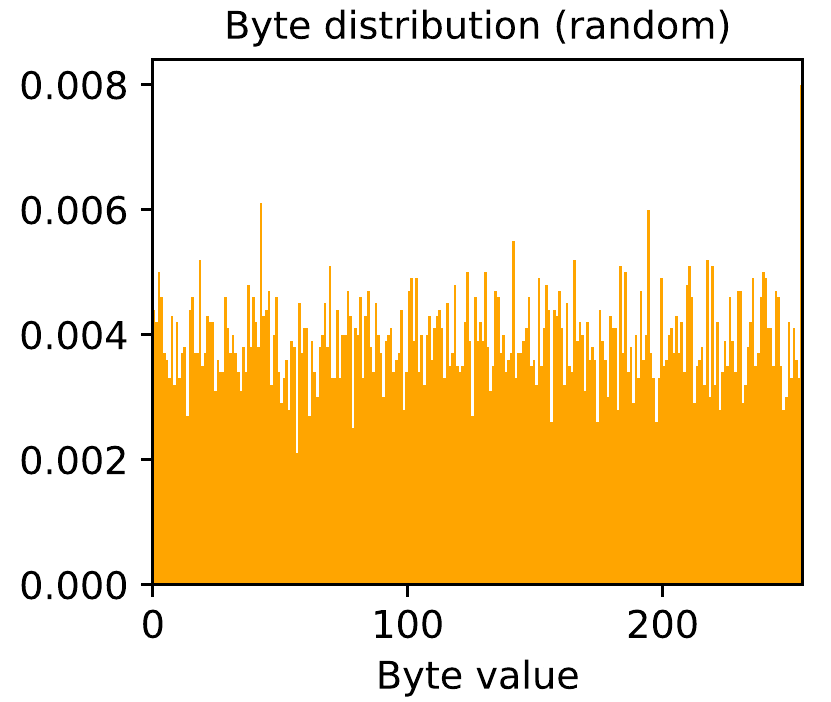}
        \includegraphics[width=0.24\textwidth]{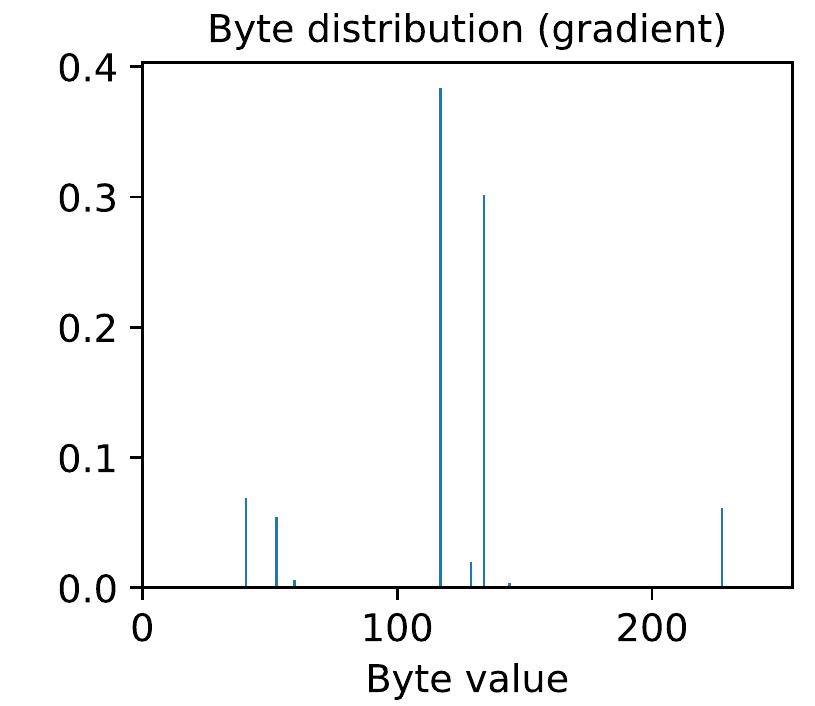}
    \caption{Distribution of the $10,000$ padding byte values injected by the random (\emph{left}) and gradient-based (\emph{right}) attacks into a randomly-picked malware sample.}
\label{fig:byte_graph}
\end{figure}

The success of our gradient-based approach relies on the fact that it guides the decision of which bytes to add, thus creating an organized \emph{padding byte pattern} specific to each sample.
To better clarify this concept, in Fig.~\ref{fig:byte_graph} we consider a sample that successfully evaded the network, and show the distribution of the $10,000$ bytes added by the two attacks. Note how, in the optimized case, only a small group of byte values is consistently injected. This shows that the gradient guides the choice of specific byte values that are repeatedly injected, identifying a clear \emph{padding byte pattern} for evasion.  

\subsection{Limitations of Our Analysis}

\begin{figure}
	\centering
	\includegraphics[width=0.42\textwidth]{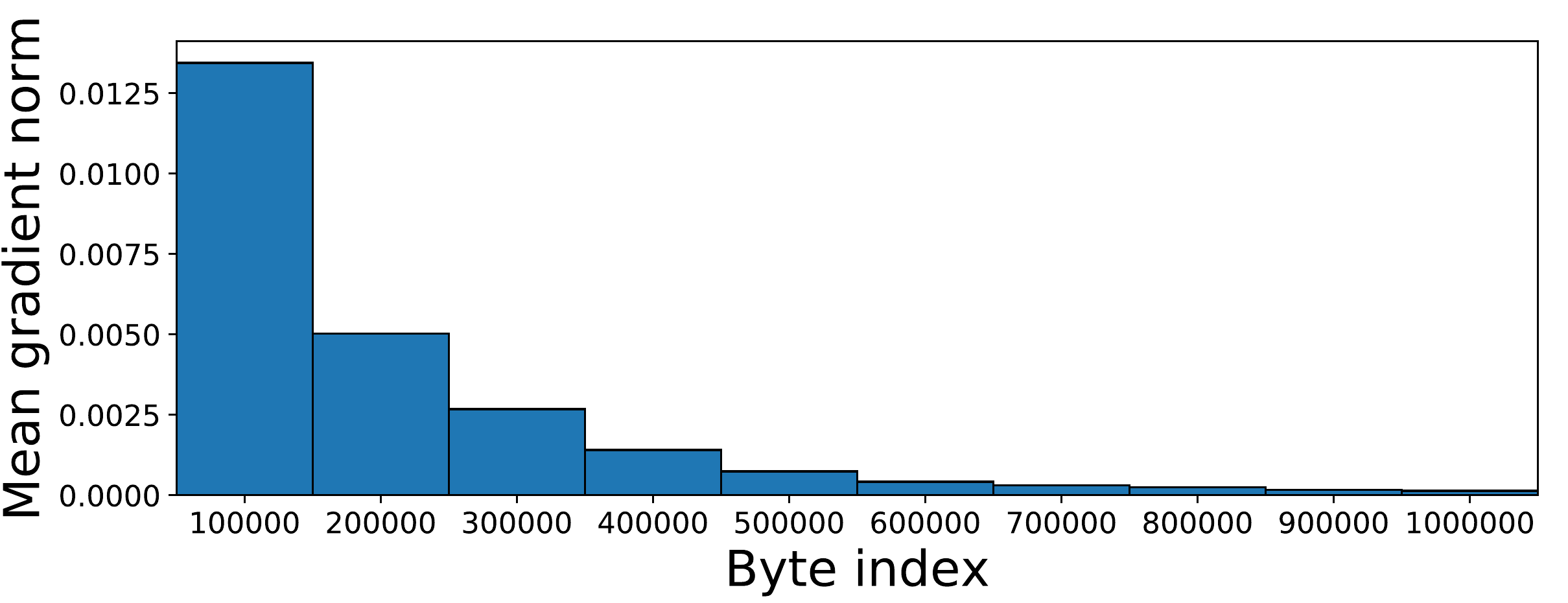}
	\caption{Mean gradient norm (per byte) over all attack samples.}
	\label{fig:gradient_norm}
\end{figure}

We discuss here some limitations related to our analysis. First, in comparison to~\cite{raff17-arxiv}, we employed a smaller dataset, and we considered an input file size $d$ of $10^6$ rather than $2 \cdot 10^6$.
These are both factors that may facilitate evasion of \emph{MalConv}.
Conversely, we found that appending bytes to the end of the file reduces the effectiveness of the gradient-based approach.
To better realize this, in Fig.~\ref{fig:gradient_norm} we show that the average norm of the gradient $\vct w$ computed over all attack samples is much higher for the \emph{first bytes} in the file. This is reasonable, as files have different lengths, and the probability of finding informative (non-padding) bytes for discriminating malware and benign files decreases as we move away from the first bytes.
From the attacker's perspective, this also means that modifying the first bytes may cause a much larger decrease of $f(\vct x)$ and, consequently, a much higher probability of evasion. 
However, as described in Sect.~\ref{sec:basics}, modifying bytes within the file may be quite complex, depending on the specific file and the content of its sections. This is definitely an interesting avenue for future research in this area.

\section{Conclusions and Future Work} 
\label{sec:conclusions}

In this work, we evaluated the robustness of neural network-based malware detection methods that use raw bytes as input. We proposed a general gradient-based approach that chooses which bytes should be modified in order to change the classifier decision. We applied it by injecting a small number of optimized bytes at the end of a set of malicious samples, and we used them to attack the \emph{MalConv} network architecture, attaining a maximum evasion rate of $60\%$. 

These results question the adequateness of byte-based analysis from an adversarial perspective. In particular, the use of deep learning on raw byte sequences may give rise to novel security vulnerabilities.  
Binary-based approaches are usually based on the hypothesis that all sections have the same importance from the \emph{learning perspective}. However, such claim is challenged by the fact that there are typically strong  semantic differences between sections containing instructions (\eg, \texttt{text}) and those containing, for example, debug information. Hence, performing manipulations directly on the targeted files might be easier than expected.

In future work, we plan to particularly investigate this issue, by exploring fine-grained, automatic changes to executables that may be more difficult to counter than the injection of padding bytes at the end of file. We also plan to repeat the assessment of this paper on a larger dataset, more representative of recent malware trends (as advocated by Rossow~\etal~\cite{rossow2012prudent}).
We anyway believe that our work highlights a severe vulnerability of deep learning-based malware detectors trained on raw bytes, highlighting the need for developing more robust and principled detection methods.
Notably, recent research on the interpretability of machine-learning algorithms may also offer interesting insights towards this goal~\cite{lipton16,kim17-arxiv}.

\end{document}